\documentclass{aa}
\usepackage{graphicx}
\usepackage{amssymb}

\usepackage{txfonts}
\usepackage{color}
\usepackage{natbib}
\usepackage{tabularx}
\usepackage{multirow}

\newcommand{\HESSJ}{HESS~J1507--622}

\begin{document}

\title{Exploring the nature of the unidentified VHE~gamma-ray~source
  \HESSJ}

\author{Wilfried Domainko\inst{1} and Stefan Ohm\inst{2,3}}

\institute{Max-Planck-Institut f\"ur Kernphysik, P.O. Box 103980, D 69029
Heidelberg, Germany
\and
Department of Physics \& Astronomy, University of Leicester, UK
\and
School of Physics \& Astronomy, University of Leeds, Leeds LS2 9JT, UK}

\offprints{\email{wilfried.domainko@mpi-hd.mpg.de}}

\date{}
 
\abstract 
{Several extended sources of very-high-energy (VHE, E $>$ 100 GeV)
  gamma rays have been found that lack counterparts belonging
    to an established class of VHE gamma-ray emitters.}
{The nature of the first unidentified VHE gamma-ray source with
  significant angular offset from the Galactic plane of 3.5$^\circ$, 
\HESSJ, is explored.}
{\emph{Fermi}-LAT data in the high-energy (HE, 100 MeV $<$ E $<$ 100
  GeV) gamma-ray range collected over 34 month are used to describe
  the spectral energy distribution (SED) of the source. Additionally,
  implications of the off-plane location of the source for a leptonic
  and hadronic gamma-ray emission model are investigated.}
{\HESSJ\ is detected in the \emph{Fermi} energy range and its
    spectrum is best described by a power law in energy with
    $\Gamma=1.7 \pm 0.1_{\mathrm{stat}} \pm 0.2_{\mathrm{sys}}$ and
    integral flux between $(0.3-300)$\,GeV of $F = (2.0 \pm
    0.5_{\mathrm{stat}} \pm 1.0_{\mathrm{sys}}) \times
    10^{-9}$\,cm$^{-2}$\,s$^{-1}$. The SED constructed from the
  \emph{Fermi} and H.E.S.S. data for this source does not support a smooth
  power-law continuation from the VHE to the HE gamma-ray range. With
  the available data it is not possible to discriminate between a
  hadronic and a leptonic scenario for \HESSJ. The location and
  compactness of the source indicate a considerable physical offset
  from the Galactic plane for this object. In case of a multiple-kpc
distance, this challenges a pulsar
  wind nebula (PWN) origin for \HESSJ\ 
since the time of travel for a pulsar born in the Galactic disk
  to reach such a location would exceed the inverse Compton (IC)
  cooling time of electrons that are energetic enough to produce
  VHE gamma-rays. However, an origin of
    this gamma-ray source connected to a pulsar that was born off the
    Galactic plane in the explosion of a hypervelocity star cannot be
    excluded.}
{The nature of \HESSJ\ is still unknown to date, and a PWN
scenario cannot be ruled out in general.
On the contrary
  \HESSJ\ could be the first discovered representative of a population
  of spatially extended VHE gamma-ray emitters with HE gamma-ray
  counterpart that are located at considerable offsets from the
  Galactic plane. Future surveys in the VHE gamma-ray range are
  necessary to probe the presence or absence of such a source
  population.}

\keywords{Gamma rays: ISM -- Stars: pulsars -- ISM: supernova remnants
  -- Radiation mechanisms: non-thermal}

\authorrunning{Domainko \& Ohm}

\titlerunning{Exploring the nature of \HESSJ}

\maketitle


\section{Introduction}

The inner Galaxy has been surveyed by the H.E.S.S. telescope array in
very-high-energy (VHE, E$>$ 100 GeV) gamma rays and during this survey
several sources of such radiation have been found which lack obvious
counterparts in other wavelength \citep{aharonian2008}. VHE gamma rays
are in many cases expected to be produced either by inverse Compton
(IC) scattering of low energy photons from e.g. stellar radiation
fields or the cosmic microwave background radiation by VHE electrons
(leptonic channel) or by $\pi^0$ decay following inelastic collisions
between hadronic cosmic rays and target protons and nuclei from the
ambient medium \citep[hadronic channel, see][for a
review]{hinton2009}. Both channels could in principle explain the
characteristics of the significant population of unidentified VHE
gamma-ray sources which lack obvious counterparts in other
wavelength. Evolved pulsar wind nebulae (PWN) as leptonic sources
\citep{dejager2008,mattana2009} or old supernova remnants (SNRs) as
hadronic sources \citep{yamazaki2006} have been proposed to be viable
object candidates. Both scenarios have in common that they assume a
rather large object age of $>10^4$ years. Indeed, as in the case of
the formerly unidentified sources HESS~J1616--508
\citep{aharonian2005} and HESS~J1731--347 \citep{aharonian2008}, both
could be associated to an evolved PWN \citep{landi2007} and an old SNR
system \citep{tian2008,abramowski2011a}, respectively. However, the
gamma-ray production processes in a large fraction of unidentified VHE
gamma-ray sources are still not known to date.

One possibility to distinguish between a leptonic and a hadronic
scenario for a specific source is to utilise observations in the to
the VHE gamma-ray range adjacent band of high-energy (HE, 100 MeV $<$
E $<$ 100 GeV) gamma rays \citep[see
e.g.][]{tam2010,abdo2011a}. \citet{torres2011} demonstrated the
potential of this approach by using data obtained in observations with
the Large Area Telescope (LAT) aboard the \emph{Fermi} satellite to
investigate the nature of the unidentified VHE gamma-ray source
HESS~J1858+020.

In the HE gamma-ray band, observations of objects located close to the
Galactic plane are severely affected by diffuse gamma-ray emission
originating from interactions of HE electrons and hadrons with
interstellar radiation fields and interstellar matter. Thus strong,
unidentified VHE gamma-ray sources with an offset from the Galactic
plane are ideal for such an investigation of the radiation mechanism.
Recently, such an object with an angular offset from the galactic
plane has indeed been found: \HESSJ\ \citep{acero2011}. This source is
detected with an integral flux above 1\,TeV of
$F_\gamma(>1\,\mathrm{TeV}) = (1.5 \pm 0.4_{\mathrm{stat}} \pm
0.3_{\mathrm{sys}} ) \times 10^{-12}$\,cm$^{-2}$\,s$^{-1}$ and a
power-law spectrum $dN/dE = k (E/1 \mathrm{TeV})^{-\Gamma}$ with
spectral index $\Gamma = 2.24 \pm 0.16_{\mathrm{stat}} \pm
0.20_{\mathrm{sys}}$ and a flux normalization of $k = (1.8 \pm 0.4)
\times 10^{-12}$\,TeV$^{-1}$\,cm$^{-2}$\,s$^{-1}$. For this particular
object, \citet{domainko2011} showed that it is difficult to explain
the gamma-ray emission to originate from an old SNR. An attempt was
made to explain \HESSJ\ in the framework of an ancient PWNe scenario
\citep{tibolla2011}. So far no pulsar could be identified in the
vicinity of the H.E.S.S. source.

Recently a GeV counterpart to \HESSJ, namely 2FGL~J1507.0--6223, has
been reported in the \emph{Fermi} 2-year catalogue
\citep{abdo2011b}. In this paper an extended 34 month \emph{Fermi}-LAT
data set is used to investigate gamma-ray emission models of \HESSJ\
using the differential energy spectrum from GeV to TeV
energies. Additionally, radio and X-ray measurements from the literature
\citep{acero2011,murphy2007} are adopted to further constrain the SED. 
Furthermore, its location offset from the Galactic plane is
used to set constraints on the nature of the source.

\section{\emph{Fermi}-LAT analysis}

The \emph{Fermi}-LAT is a pair-conversion telescope operating in the
20\,MeV to 300\,GeV energy range. The direction of incident photons is
reconstructed by means of a tracker. A calorimeter is used to
measure the energy of the incident particle, and an anti-coincidence
system efficiently suppressed the charged-particle background. The
large field-of-view of the instrument of $\sim 2.4$\,sr allows to
cover the whole sky every two orbits. The LAT provides an angular
resolution of $<1^\circ$ at 1\,GeV and $<0^\circ.2$ at 10\,GeV. A full
mission- and instrument-related description can be found in
\citet{Fermi:Inst}.

Based on 24 months of data, the HE gamma-ray source 2FGL~J1507.0--6223
has been reported to emit gamma-ray emission in the HE regime
\citep{abdo2011b}. Within the $1\,\sigma$ positional uncertainty,
2FGL~J1507.0--6223 is consistent with the best fit position of
\HESSJ\ (angular separation of $3.6'$) which makes a physical
association of the \emph{Fermi}-LAT and the H.E.S.S. source very
likely. The best fit model to the spectral energy distribution (SED)
of 2FGL~J1507.0-622 reported in \citet{abdo2011b} is a power law in
energy ($dN/dE \propto E^{-\Gamma}$) with differential index
$\Gamma=1.9\pm0.2$ and an integral flux between ($1-100$)\,GeV of
$(1.0\pm0.3)\times 10^{-9}$\,cm$^{-2}$\,s$^{-1}$. Note that in the
catalogue significant emission is reported only in the ($10-100$)\,GeV
band and flux upper limits are given at lower energies.

The analysed data set used in this work has been recorded in
observations over 34 months of operation from the commissioning of
\emph{Fermi} in August 2008 until May 2011. Data analysis has been
performed using events with reconstructed energies between 300\,MeV
and 300\,GeV, utilising the Fermi Science Tools (FST) package, version
v9r23p1\footnote{http://fermi.gsfc.nasa.gov/ssc/data/analysis/documentation/Cicerone/}. Events
have been selected according to and analysed with the
\emph{P7Source\_V6} instrument response functions (IRFs).

The best fit position of \HESSJ\ has been used to define a
Region-of-Interest (RoI) of $10^\circ$ from which events have been
included in the maximum likelihood analysis. All sources within
$15^\circ$ around the RoI were modeled to produce energy spectra and
to calculate the test statistics (TS) \citep{Mattox1996} of the
source. With the FST $gtlike$ tool and an unbinned maximum likelihood
fitting procedure, the energy spectrum of 2FGL~J1507.0-622 has been
derived. For this purpose, all sources listed in the \emph{Fermi}-LAT
2-year catalogue in the RoI were modeled according to their best fit
model. In order to guarantee stable fit results and since there are 65
identified \emph{Fermi}-LAT sources in this region of the sky, only
the flux at the de-correlation energy has been left as free parameter
in the minimisation procedure. All other parameters have been fixed to
their catalogue value. The Galactic and Extragalactic background
components were modeled with the files $gal\_2yearp7v6\_v0.fits$ and
$iso\_p7v6source.txt$\footnote{http://fermi.gsfc.nasa.gov/ssc/data/access/lat/BackgroundModels.html},
respectively. At the \HESSJ\ position a TS value of 60,
  corresponding to a statistical significance of $\sim 7.7\sigma$ is
  found. The energy spectrum is best described by a power law in
  energy with $\Gamma=1.7 \pm 0.1_{\mathrm{stat}} \pm
  0.2_{\mathrm{sys}}$ and integral flux between $(0.3-300)$\,GeV of $F
  = (2.0 \pm 0.5_{\mathrm{stat}} \pm 1.0_{\mathrm{sys}}) \times
  10^{-9}$\,cm$^{-2}$\,s$^{-1}$. The flux in each energy band as shown
  in Fig.~\ref{figure:SED_lep} and Fig.~\ref{figure:SED_had} has been
  calculated following the same procedure as described above.

\begin{figure}[ht]
\centering
\includegraphics[height=7cm]{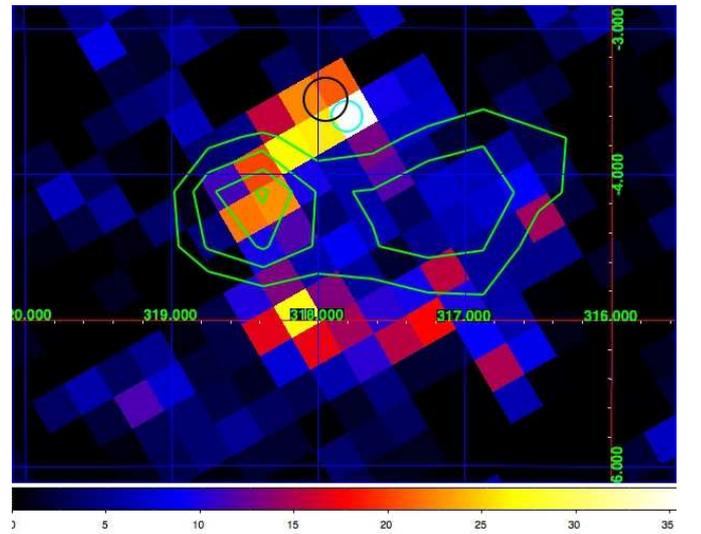}
\caption{Test statistics map for photon energies $10\leq E \leq
  300$\,GeV. The cyan circle denotes the error circle of the best fit
  position of the LAT source in this energy band and the black circle
  indicates the intrinsic width of the H.E.S.S. source. Overlaid in
  green are CO contours \citep{Dame01} of the foreground molecular
  cloud located at a distance of $\approx 400$\,pc (-10 to
  0\,km\,s$^{-1}$) and measured in units of average temperature (from
  0.75 to 1.5\,K in steps of 0.25\,K).}
\label{figure:skymap}
\end{figure}

The systematic error of the spectral result has been estimated by
  employing alternative source models, where other spectral parameters
  such as the spectral index for power-law-type sources and/or the
  normalisation of 2FGL sources within 5$^\circ$ of 2FGL~J1507.0-622
  have been left free in the minimisation procedure. As can be seen in
  Fig.~\ref{figure:skymap}, there is some residual emission above
  10\,GeV energies, not covered by the Galactic diffuse model and
  partly coincident with a molecular cloud located at $\sim400$\,pc
  distance \citep{Dame01,acero2011}. Given the distance of
  2FGL~J1507.0-622 to the CO peak of $\sim 1^\circ$, no significant
  contribution to the 2FGL~J1507.0-622 emission is expected. Indeed,
  including an additional point-like source at the peak CO emission in
  the model did not change the fit results by more than 10\% above
  1\,GeV (20\% below 1\,GeV).

\section{Spectral energy distribution}
\label{sec:sed}

\begin{figure*}[ht]
\centering
\includegraphics[height=8cm]{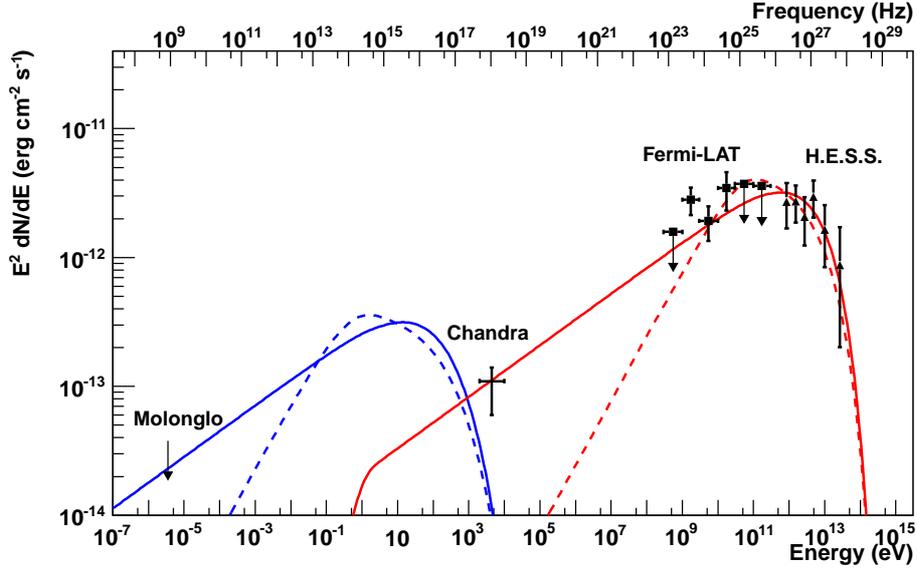}
\caption{Observed and calculated leptonic SED for \HESSJ\ for a continuous injection of electrons.
Two cases are shown:
solid lines represent a model with an assumed (young) objects age of $10^4$~years and 
rather steep electron injection spectrum of $\Gamma_\mathrm{e}=2.6$; the dashed lines
represent a rather old object with an age of $3 \times 10^5$~years
and an electron injection spectrum with $\Gamma_\mathrm{e}=2.0$.
For more details on the modeled SED see main text.}
\label{figure:SED_lep}
\end{figure*}

The \emph{Fermi} results presented in the previous section together
with H.E.S.S. measurements are now used to study the SED of \HESSJ.
Additionally, \citet{acero2011} found diffuse X-ray emission using 
\textit{Chandra} within the extend of \HESSJ\ \citep[magenta circle in
  Fig. 3 in][]{acero2011}. This source features a flux of
  $(1.1^{+0.3}_{-0.5}) \times 10^{-13}$\,erg\,cm$^{-2}$\,s$^{-1}$ in
  the energy band of (2 -- 10)~keV. This is the only extended X-ray
  source inside the VHE gamma-ray source and thus it is a potential
  counterpart to \HESSJ. However, it has to be noted that a firm
  association between the diffuse X-ray source and the VHE gamma-ray
  source can not be established \citep[see][]{acero2011}. To
  investigate possible constraints on the magnetic field, here it is
assumed that this extended X-ray source is a counterpart to
\HESSJ. Finally, the position of this source has also been covered
by the second epoch Molonglo Galactic Plane Survey in radio at 843~MHz \citep{murphy2007}. 
No source has been detected at the position of \HESSJ\ within the sensitivity limit
of this survey of 10~mJy/beam (beam size: $45 \times 45|\delta|$~arcsec). 
The extend of \HESSJ\ contains about 450 beams. If the sensitivity per
beam is integrated over the whole area of the gamma-ray source then
the limit on the radio flux from \HESSJ\ would be conservatively 4.5 Jy.

The observed SED does not strongly support a smooth power-law
continuation with slope $\Gamma = 2.24$ from \HESSJ\ to the HE regime
(see Fig \ref{figure:SED_lep}) and is suggestive of spectral
curvature. It is apparently compatible with a flat spectrum with
$\Gamma \simeq 2$ but also a rising slope from the HE to the VHE
gamma-ray energy range cannot be excluded. With the available data it
is not possible to discriminate between a hadronic and a leptonic
scenario for \HESSJ\ based on the spectral form (see
e.g. Fig. \ref{figure:SED_lep},\ref{figure:SED_had}) and hence both
scenarios are discussed in the following.

\subsection{Leptonic model}

In Figure \ref{figure:SED_lep}, a computed SED for the simple case of a
one zone model \citep[see][]{hinton2007} is
compared to the observational data. Given the fact that very little is
known about \HESSJ, this simple approach was chosen in order to avoid
too many implicit assumptions on the nature of the source. Due to the
off-plane location of \HESSJ\ it is furthermore reasonable to assume
that only photons from the cosmic-microwave background (CMB) are
up-scattered to VHE energies by HE electrons and that contributions
from other target radiation fields (e.g. starlight) can be
neglected. In order not to overproduce synchrotron emission in the
X-ray band in this scenario, the strength of the magnetic field was
assumed to be $B = 1\,\mu$G. 
As a first approach a quite young age of $10^4$ years of the source
is assumed and electrons are continuously injected during this period.
In this case the rather flat observed SED from the HE
to the VHE gamma-ray energy range requires a rather steep spectral
index of the injected electrons. Fig. \ref{figure:SED_lep}
demonstrates that such a leptonic scenario can in principal reproduce
the observational data (input parameters: electron spectra
  following a power law with exponential cut-off with index
  $\Gamma_\mathrm{e}=2.6$ and cut-off energy $E_{c,e} = 60$\,TeV, total energy in
  electrons $E_e = 2.5 \times 10^{49}\,(d/ 1 \mathrm{kpc})^2$\,erg, a
  magnetic field of $B = 1\,\mu$G, and age of $10^4$~years). The constraints on the magnetic
  field strength inferred here differ from the value estimated in
  \citet{acero2011} of $B = 0.5\,\mu$G due to the different spectral
  shape of the considered electron population \citep[$\Gamma_\mathrm{e}=2.2$
  in][]{acero2011}. 
This leptonic model would furthermore predict a radio flux density
of about 2~Jy at the Molonglo frequency of 843~MHz. That would be comparable to
the radio limit on the entire area of the gamma-ray source.
As a further constraint, this model would require a quite low
density of the ambient medium of $\lesssim 0.1$~cm$^{-3}$. For higher densities
bremsstrahlung would start to become dominant below a few GeV energies 
for such a steep electron spectrum.
It has to be noted that, since the SED of \HESSJ\
is observationally not very well constrained the parameters of this
leptonic model are not unique to describe the observations (see
e.g. Fig. \ref{figure:SED_lep},\ref{figure:SED_had}).

The population of electrons required to explain the gamma-ray
  emission in a leptonic scenario could in principle be generated by
  the spin-down energy loss of a pulsar. In the case where the pulsar
  is significantly older than its characteristic spin-down time the
  evolution of the population of electrons can be approximated by a
  burst like injection followed by IC and synchrotron cooling. For a
  review on the evolution of PWNe see \citet{gaensler2006}. Evolved
  PWNe  with an age $\gg 10^{4}$~years can appear as rather bright VHE
  sources with only faint multi-wavelength counterparts
  \citep{dejager2009}.

As a second possibility for a leptonic model that exhibits a flat
gamma-ray spectrum from the HE to the VHE gamma-ray energy band 
a model is chosen where electrons with a hard spectral index 
are continuously injected 
over a quite long period of time and
interact with target magnetic and radiation fields
\citep{hinton2009} (input parameters: electron spectra
  following a power law with exponential cut-off with index
  $\Gamma_\mathrm{e}=2.0$ and cut-off energy $E_{c,e} = 60$\,TeV, total energy in
  electrons $E_e = 2 \times 10^{47}\,(d/ 1 \mathrm{kpc})^2$\,erg, a
  magnetic field of $B = 1\,\mu$G). 
With these parameters it is found that an age of the source 
of $3 \times 10^5$~years would be required to explain the gamma-ray
emission in the GeV range (see Fig. \ref{figure:SED_lep}).
Thus \HESSJ\ could in principle also be
explained by a rather old object containing a cooling population of electrons
that is continuously injected.

\subsection{Hadronic model}

\begin{figure*}[ht]
\centering
\includegraphics[height=8cm]{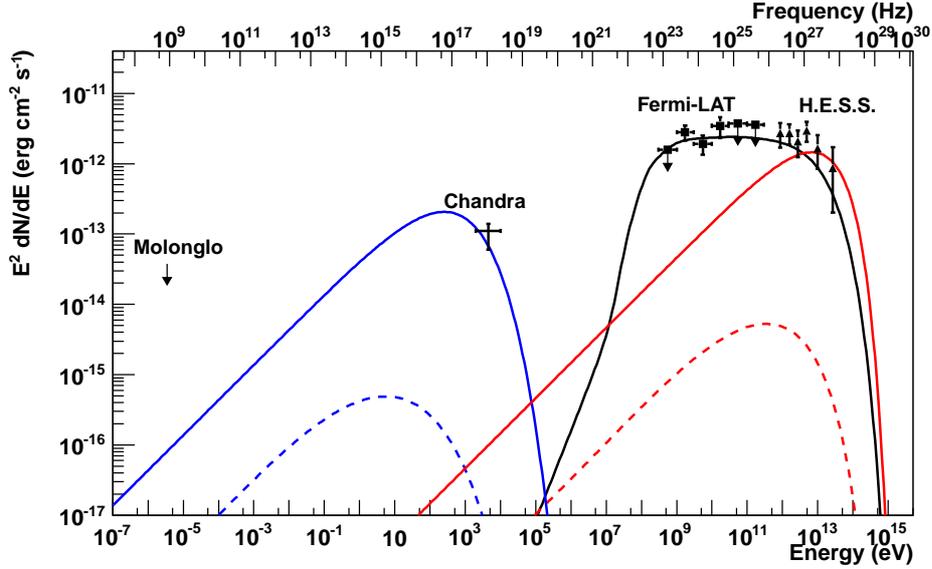}
\caption{Observed and calculated hadronic SED for \HESSJ\ for a continuous
injection of protons. In blue and red synchrotron and IC
radiation of primary (solid lines) and secondary (dashed lines) electrons is shown for a source age of $10^4$~years. For
  more details see main text.}
\label{figure:SED_had}
\end{figure*}

In Figure \ref{figure:SED_had}, a computed hadronically dominated model is compared
to the observed SED. For a comprehensive discussion of this hadronic
model see \citet{domainko2011}. The input parameters for this model
are: total energy in hadronic cosmic rays $E_{CR} =
10^{50}$~erg, spectral index of injected protons of $\Gamma = 2.0$, cutoff energy
$E_{c,CR} = 100$\,TeV, distance $d=2$\,kpc, density of the ambient
medium $n_{\mathrm{H}} = 1$\,cm$^{-3}$, electron to proton ratio
$\epsilon = 5 \times 10^{-4}$, a magnetic field of $B$ = 1$\mu$G
and an age of $10^4$~years.
The predicted radio flux density at 843~MHz of this model
is 5~mJy, compatible with the non-detection in the Molonglo survey.
This scenario would require a rather young SNR and such a
model has been regarded as an unlikely explanation for \HESSJ\
\citep{domainko2011}. This conclusion is based on the compactness of
the VHE source measured in \citet{acero2011}.
As can be seen from Fig. \ref{figure:SED_had} the IC and synchrotron contribution
from secondary electrons for an object age of $10^4$~years is orders of magnitude
below the IC and synchrotron emission from primary electrons.
Note that only for object ages of $\gtrsim 10^6$~years emission from secondary electrons become
important but even for this case they are still compatible with the radio limits and the \emph{Chandra}
flux point.

A second possibility for a hadronic scenario is given by the location
on the sky of \HESSJ\ in the direction of the outskirts of a molecular
cloud \citep[see][]{acero2011}. In this scenario a locally enhanced
cosmic ray density at the source location could result in detectable hadronic
gamma-ray production. However, it has to be noted that in
\citet{acero2011} it has been concluded that the substantial
difference in extension and an offset of $\sim1^\circ$ of the VHE
source centroid from the densest part of the molecular cloud suggests
no obvious association between these two objects
(see also Fig. \ref{figure:skymap}). The molecular cloud
is most likely located at a distance of about 400\,pc. Therefore, a
total energy in hadronic cosmic rays of $E_{CR} \approx 2.4 \times
10^{47}\,(n/ 10\,\mathrm{cm}^{-3})$\,erg ($n$: density of target
material) would be required to explain the gamma-ray emission. This
population of cosmic rays would have to be concentrated in a sphere
with radius of about 1\,pc (corresponding to an angular radius of
0.15$^\circ$ at 400\,pc). Consequently, the energy density in cosmic
rays at \HESSJ\ for an assumed density of target material of $n
\approx 10$\,cm$^{-3}$ would have to exceed the local cosmic ray
density at the solar system by about three orders of magnitude. The
absence of any plausible particle accelerator in the vicinity of
\HESSJ\ points against such a possibility and thus also this scenario
may not provide a convincing explanation for the nature of the source.

In contrast to these hadronic scenarios here in the following a
leptonic model for \HESSJ\ is considered.

\section{Discussion}

In this section the implications of a leptonic model in general and an
evolved PWN origin in particular for \HESSJ\ are discussed. This
scenario can explain the appearance of the source as relatively bright
VHE gamma-ray emitter with only faint low-energy counterpart
\citep{dejager2008,mattana2009}.

\subsection{Estimation of the distance}
\label{sec:distance}

The distance to an extended unidentified source can in principle be
estimated if its physical size can be constrained. VHE particles
escape from their production site by diffusion and for the simple case
of point-like injection the size of a source is given by the diffusion
coefficient and the age of the particles. For the case of old leptonic
sources, the size of the object is limited by energy losses of the VHE
electrons and thus the age of these particles is equivalent to the
electron cooling time. At locations far away from the galactic plane
the dominating photon field responsible for energy losses of the VHE
electrons is the CMB with an energy density of about
0.3\,eV\,cm$^{-3}$. For this case the cooling time of electrons is
given by $t_\mathrm{cool} \approx 10^6 \, (E_\mathrm{e}/ 1 \,
\mathrm{TeV})$ years \citep{hinton2009} with $E_\mathrm{e}$ being the
electron energy. The slowest possible diffusion is Bohm diffusion with
a diffusion coefficient of $ \kappa_\mathrm{Bohm} \approx 10^{26} \,
(E_\mathrm{e, TeV})/(B_\mathrm{\mu G})$ cm$^2$ s$^{-1}$ where
$E_\mathrm{e, TeV}$ gives the electron energy in TeV and
$B_\mathrm{\mu G}$ gives the magnetic field in $\mathrm{\mu G}$. With
these assumptions the physical extension of a cooling dominated
leptonic source can be estimated by $r \approx \sqrt{2 \,
  \kappa_\mathrm{Bohm} \, t_\mathrm{cool}}$.  For electrons with an
energy of 60 TeV and a magnetic field of 1$\mu$G corresponding to $
\kappa_\mathrm{Bohm} \approx 6 \times 10^{27}$ cm$^2$ s$^{-1}$ and
$t_\mathrm{cool} \approx 2 \times 10^4$ years the source would have a
radius of about 25 pc. This estimate is also comparable to the typical
sizes of evolved PWN which are abundant extended VHE gamma-ray sources
in the galactic plane \citep[e.g.][]{mattana2009,kargaltsev2010}. If
\HESSJ\ has indeed an extension of 25 pc its angular radius of
0.15$^\circ$ would imply a distance of about 10 kpc. It has to be
noted that a larger magnetic field would result in slower diffusion
and thus in a more compact source morphology. However, the low X-ray
flux in comparison to the gamma-ray flux does not support such a
possibility (see Fig. \ref{figure:SED_lep}).

The more complete multi-wavelength picture of \HESSJ\ might place
further constraints on its distance. The SED fit from
Sec.~\ref{sec:sed} seems to indicate a quite low magnetic field of
$\sim 1\, \mu\mathrm{G}$ which is significantly smaller than the
typical Galactic magnetic field of $3 - 10\,\mu\mathrm{G}$. One
possibility for the source being situated in a region with low
magnetic field is a location far below the plane and thus outside the
magnetic field of the Galactic disk. \citet{simard1980} found a
thickness of the magneto-ionic plane of the Milky Way of about 1.4
kpc. Thus, if \HESSJ\ would be hosted by a low magnetic field
environment outside this plane, its distance would exceed 20 kpc.
Such a situation would imply then a physical radius of the object of
$\gtrsim50$ pc. For comparison, this value corresponds roughly to the
extension of the largest VHE gamma-ray emitting PWN found so far
\citep{kargaltsev2010}.

\subsection{Energetics of the source}

The flux of \HESSJ\ corresponds to a luminosity of
$L_\gamma(>1\mathrm{TeV}) \approx 8 \times 10^{32} (d / 1 \,
\mathrm{kpc})^2$ erg s$^{-1}$ with $d$ being the distance to the source
in kpc. For the distance estimates from Sec.~\ref{sec:distance} of
10\,kpc (20\,kpc) this would result in a VHE gamma-ray luminosity of
about $8 \times 10^{34}$\,erg\,s$^{-1}$ ($3 \times
10^{35}$\,erg\,s$^{-1}$) which is again comparable to the most
luminous PWN detected so far
\citep[e.g.][]{mattana2009,kargaltsev2010}.

To estimate the total energy in electrons required to explain the
luminosity of the gamma-ray source, two different approaches are
considered here.

Firstly, only the population of electrons which can directly be probed
by the gamma-ray observations are considered. As has been shown in
Sec.~\ref{sec:sed} gamma-ray emission from \HESSJ\ has been detected
down to the (1~--~3)~GeV band which requires electrons with energies of
about 100~GeV for IC scattering on the CMB. These electrons IC cool on
a timescale of $t_\mathrm{cool} \approx 10^7$\,years. Since the total
flux in the \emph{Fermi} energy range is
$F_\gamma \approx 8.2 \times 10^{-12}$\,erg\,cm$^{-2}$\,s$^{-1}$ the
total energy in electrons above 100\,GeV, given by $E_\mathrm{tot}
\approx F_\gamma t_\mathrm{cool} 4 \pi d^2$, amounts to $3 \times
10^{47} (d/1\,\mathrm{kpc})^2$\,erg with $d$ being the distance to the
source.

Secondly, for determining the total energy in electrons the entire
energy range for the electrons from the sub-relativistic regime up to
the cutoff at 60\,TeV is considered. It is further assumed that the
electron distribution over the whole energy range can be described by
a power-law distribution with index 2.6. From the calculated SED it is
found that a total energy in electrons for this case of about $2.5
\times 10^{49}\, (d/ 1 \mathrm{kpc})^2$\,erg would be needed to explain
the luminosity of the source.

With the estimates from the previous paragraphs the energetics in
electrons for the putative source distances can be determined.  A
distance of 10 kpc (20 kpc) would require a total energy in VHE
electrons of about $3 \times 10^{49}$\,erg ($1.2 \times 10^{51}$\,erg)
for the case where only the gamma-ray producing electrons are
considered or $2.5 \times 10^{50}$\,erg ($ 10^{52}$\,erg) for the case
where the entire energy range for the electrons is taken into
account. These energy requirements would be compatible with the
typical spin-down energy of a pulsar of a few times $10^{49}$\,erg
\citep{gaensler2006} for the first case up to a distance of about
10\,kpc but would exceed the typical energetics of pulsar wind nebulae
for the second case. For the second case a pulsar can only provide the
necessary energy in electrons if the source is located within a
distance of a few kpc. The compact morphology of \HESSJ\ seems to
challenge this scenario (see Sec.~\ref{sec:distance}).

\subsection{PWN interpretation and potential Galactic halo location}
\label{sec:halo}

The rather compact appearance of \HESSJ\ favors a multi-kpc distance
$d$ of the source (see Sec.~\ref{sec:distance}) which in turn results
in a considerable physical offset from the Galactic plane of about
$600\,(d/10\,\mathrm{kpc})$\,pc. Such a location has certain
implications for a PWN scenario for the VHE gamma-ray source since
pulsars are generally born in the Galactic disk. However, pulsars can
be ejected from their birth place due to natal kicks that results in a
Galactic lateral distribution for young pulsars which follows a
Gaussian distribution with age dependent scale height. The width of
this distribution scales linearly with pulsar age and for pulsar ages
$\lesssim 1$\,Myr it is expected to be $\lesssim 500$\,pc
\citep{sun2004}. Consequently, gamma-ray PWNe hosting pulsars with
ages typically shorter than a few times 10$^5$\,years are expected to
be found at offsets considerably smaller than 500\,pc and thus at
smaller offsets than has been estimated for \HESSJ. Even for extreme
cases of pulsar velocities of 1000\,km/s \citep{hobbs2005} the pulsar
would need $ 6\times 10^5$\,yr to travel 600\,pc. 
A pulsar located at such a
large distance from the plane would be older than the typical
gamma-ray PWN hosting pulsar and its age would have exceeded the IC
cooling time of electrons that are energetic enough to produce
VHE gamma-rays (see \ref{sec:distance}). Thus, the off-plane location of
\HESSJ\ challenges a PWN origin for this source.

This limitation on a PWN origin can in principle be overcome by a
scenario where the pulsar is born offset from the Galactic disk in the
core-collapse of a massive star that was ejected from the Galaxy
\citep[see][for core-collapse supernovae connected to hypervelocity
stars]{zinn2011}. Since the pre-supernova lifetime of massive stars is
of the order of $10^{7}$\,years, significantly longer than the
lifetime of young energetic pulsars, such a star can reach larger
offsets from the Galactic plane. Consequently, after the star
  has exploded as a supernova, an energetic pulsar could be found at
  such an off-plane location. For this scenario it would be expected
that a SNR shell should be found in the neighbourhood of the PWN.
Therefore one restriction for such a model is the absence of a SNR
shell in the vicinity of \HESSJ. Gamma-ray PWNe could be much
  older than the duration of the Sedov phase of the SNR. In this case
  the very old PWN would be very extended. Deep future VHE gamma-ray
  observations that investigate the presence of emission with lower
  surface brightness on larger spatial scales could be used to test
  this scenario.

It has further to be noted that supernovae of hypervelocity stars
appear to be rare events \citep{zinn2011}. Therefore only a
  very limited number of PWNe resulting from such supernovae are
  expected in the Milky Way. Consequently, this scenario could also be
  tested by surveys that search for similar sources as \HESSJ\ (see
  Sec.~\ref{sec:population}).

The possibility that \HESSJ\ belongs to the Galactic halo discussed in
this section is based on the observational result that the source is
rather compact and located at significant angular distance from the
Galactic plane. However, future more sensitive VHE gamma-ray
observations could in principle reveal that the source is more
extended than measured to date. In this case, the constraints on the
distance and on the physical offset from the Galactic plane would be
relaxed and a PWN interpretation would appear more feasible. A
decreased distance to \HESSJ\ would also result in a decreased
gamma-ray luminosity and hence also a less powerful pulsar would be
required to explain the VHE gamma-ray source.

To summarise, the properties of \HESSJ\ challenge a PWN
origin but due to the uncertainty in distance such a scenario 
cannot be ruled out in general.

\subsection{Potential population of such objects}
\label{sec:population}

In this section an alternative to a PWN scenario is explored.
The following considerations are based on the result that with 
\HESSJ\ and 2FGL~J1507.0--6223 an unassociated HE source has been linked to
an unassociated VHE source \citep{abdo2011b}. It is further based on the 
discovery of a considerable number of unassociated HE sources reported 
in the \emph{Fermi} 2-year catalogue \citep{abdo2011b}. In principle
some of these HE sources could feature extended VHE emission similar
to \HESSJ.

So far only one VHE gamma-ray source without prominent
multi-wavelength counterpart, namely \HESSJ\ has been found outside
the main distribution of such sources within $\pm2^\circ$ around the
Galactic plane \citep{chaves2009}. However, \HESSJ\ could in principle
be the first discovered representative of a source population which is
distributed with large angular offset around the Galactic disk. In
this case the distribution of this object population can be used to
further constrain the distance and thus nature of their
representatives. For the simplest case, where sources from this
population are distributed homogeneously around the Earth, the surface
density of these objects should be constant in every direction. Since
the distance to the source is likely to be several kpc (see
Sec.~\ref{sec:distance}) it may belong to a population of objects
which is distributed around the Galactic center rather than to a local
source population around the Earth. In this case the surface number
density should be more concentrated around the galactic center.

\HESSJ\ is a rather bright VHE source and is observed with
about 8\% of the flux of the Crab nebula \citep{acero2011}. Therefore,
the H.E.S.S. galactic plane survey would have sufficient sensitivity
for detecting a source like \HESSJ\ out to a Galactic latitude of
$\pm4^\circ$ and the result that only one such source has been found
can be used to constrain their number on the sky. Since the
H.E.S.S. Galactic plane survey extends over 140$^\circ$ in Galactic
longitude the surface density of objects like \HESSJ\ in the bands of
-4$^\circ$ to -2$^\circ$ and 2$^\circ$ to 4$^\circ$ Galactic latitude
should not be much larger than one source per $\sim0.17$
steradians. 
For an order of magnitude estimate on the number of comparable
sources here a constant source number density is assumed as would
be expected for a local source population. In this case
$\lesssim 70$ sources would be distributed on the entire sky. 
For different distributions (e.g. concentrated on the galactic
center as discussed above) this estimate is only a course guideline. 
Nevertheless, in principle $\mathcal{O}(10)$ similar sources might be scattered 
on the sky.
Surveys obtained with
the future Cherenkov Telescope Array \citep[CTA][]{CTA} and
HAWC\footnote{http://hawc.umd.edu/} will be
crucial to test if \HESSJ\ belongs to a population of sources with
similar characteristics.

\section{Summary and outlook}

In this paper the SED from GeV to TeV energies for \HESSJ\ has been
explored and additional results from radio and X-ray observations from the 
literature have been used to further constrain the emission scenarios. 
Based on the spectral results and with the available data it is not possible to discriminate
between a hadronic and a leptonic scenario. The location and
compactness of the source may indicate a considerable physical offset
from the galactic disk for this object. Interestingly, such a scenario
would imply that this source is disconnected from a young stellar
population. So far only very few Galactic VHE gamma-ray sources that
are related to an old stellar population have been found. Such sources
include remnants of supernovae type Ia \citep{acero2010,acciari2011}
and most likely a globular cluster \citep{abramowski2011b}, but none
of them lacks a prominent low-energy emitter in its angular
vicinity. The nature of \HESSJ\ is still unknown to date, but a PWN
scenario cannot be ruled out in general. More
sensitive VHE gamma-ray observations are desirable to search for
spatially more extended, low surface brightness VHE gamma-ray emission
from this source. At the same time, further MWL observations in the
X-ray and radio band appear to be necessary to constrain the nature of
the gamma-ray emitter. These observations may also help to explore if
this source is indeed linked to an old stellar population.

\HESSJ\ could be the first representative of a class of VHE gamma-ray
sources with HE counterpart that are distributed with larger physical
offset around the Galactic disk. In this context it is important to
note that several hundred HE gamma-ray sources have been reported in
the \emph{Fermi} 2-year catalogue \citep{abdo2011b} that appear to be
unassociated with counterparts in other wavelength. In principle some
of them could be linked to extended VHE gamma-ray sources similar to
\HESSJ. Future surveys in the VHE gamma-ray range are necessary to
explore whether there is indeed a Galactic halo source population in
this energy range.

\begin{acknowledgements}
  SO acknowledges the support of the Humboldt foundation by a
  Feodor-Lynen fellowship.  The authors want to thank V. Marandon for providing the CO map
  and J. A. Hinton for valuable discussion.
  The authors acknowledge support from their
  host institutions.
\end{acknowledgements}

\end{document}